\documentclass[prd,aps,twocolumn,floats,floatfix,superscriptaddress,preprintnumbers,showpacs,nofootinbib]{revtex4-1}
\usepackage{amsfonts,amsmath,epsfig,amssymb,latexsym,eucal,color,graphicx,array,subfigure,bm,mathrsfs}
\begin{document}
\preprint{RUP-19-25}
\title{Dynamical Transition from a Naked Singularity to a Black Hole}
\author{Kenta {\sc Hioki}}
\email{Corresponding author; Hioki\_Kenta@dn.smbc.co.jp}
\address{Sumitomo Mitsui Banking Corporation, 1-2, Marunouchi 1-chome, Chiyoda-ku, Tokyo 100-0005, Japan}
\author{Tomohiro {\sc Harada}}
\email{harada@rikkyo.ac.jp}
\address{Department of Physics, Rikkyo University, Toshima, Tokyo 171-8501, Japan}
\begin{abstract}
We show that a Reissner-Nordstr\"{o}m (RN) black hole can be formed by dropping a charged thin dust shell onto a RN naked singularity.
This is in contrast to the fact that a RN naked singularity is prohibited from forming by dropping a charged thin dust shell onto a RN black hole.
This implies the strong tendency of the RN singularity to be covered by a horizon in favour of cosmic censorship.
We show that an extreme RN black hole can also be formed from a RN naked singularity by the same process in a finite advanced time.
We also discuss the evolution of the charged thin dust shells and the causal structure of the resultant spacetimes~\footnote{The statements expressed in this paper are those of the authors and do not represent the views of Sumitomo Mitsui Banking Corporation or its staff.}.
\end{abstract}
\maketitle

\section{Introduction}
\label{sec1}
The weak cosmic censorship conjecture states that any spacetime singularity caused by a gravitational collapse must be hidden inside an event horizon~\cite{Penrose:1969pc}.
This weak cosmic censorship conjecture plays  one of the important roles as a guide to establish new physics beyond general relativity.
Many attempts have been so far performed to verify the weak cosmic censorship conjecture by either gedanken experiments or numerical analyses.

In gedanken experiments, by dropping a test particle with a sufficiently large electric charge and/or angular momentum into an extreme black hole, they can study whether or not the black hole can turn into a naked singularity.

One early gedanken experiment showed that weak cosmic censorship is preserved, because a spinning test body cannot plunge into an extreme Kerr-Newman black hole, thereby preventing the formation of a naked singularity~\cite{Wald:1974}.
Related studies have been presented in~\cite{Needham:1980fb, Semiz:1990fm, deFelice:2001wj}.
It has also been shown that one can make a nearly extreme black hole jump over the extremality by tossing into it a test body composed of reasonable matter satisfying the energy conditions~\cite{Hubeny:1998ga, Jacobson:2009kt, Chirco:2010rq}.
In order to justify these results, however, we have to take into account the self-energy of a test body and its back reaction~\cite{Hod:2002pm, Barausse:2010ka, Isoyama:2011ea}.
It has also been proven that no gedanken experiments can ever succeed in forming an overcharging and/or overspinning ``black hole,'' provided only that the non-electromagnetic stress energy tensor satisfies the null energy condition~\cite{Sorce:2017dst}.
Gedanken experiments considering higher dimensions or quantum effects have also been studied~\cite{BouhmadiLopez:2010vc, Matsas:2007bj, Richartz:2008xm, Matsas:2009ww}.

Some attempts have been made to overcharge a black hole by using a collapsing charged thin shell, for which we can take the back reaction into account.
In this case, it has been found that  the self-repulsion of the thin shell prevents it from collapsing into the event horizon of the black hole~\cite{Hubeny:1998ga, Gao:2008jy}.
The evolution of a charged thin shell in a non-asymptotically flat spacetime has also been studied, and the results support the cosmic censorship conjecture~\cite{Lake:1979zzb}.

Although many gedanken experiments support the weak cosmic censorship conjecture, possible counterexamples have been found in numerical simulations.
For examples, the collapse of spheroidal configurations of collisionless particles has been considered~\cite{Shapiro:1991zza}.
Sufficiently large spheroids give rise to divergent curvature but produce no apparent horizon.
Recently, it has been found that the peak values of the curvature invariants start to decrease gradually with time ~\cite{Yoo:2016kzu}.
In another setup, time-dependent solutions of the Einstein-Maxwell equations using anti-de Sitter boundary conditions have been presented~\cite{Horowitz:2016ezu, Crisford:2017zpi}.
The electric field and hence the curvature grows as a power law in time over a vast region visible from infinity.
In a broad context, there are arguments as to whether or not this example is physical~\cite{Crisford:2017gsb, Horowitz:2019eum}.
Black holes in higher dimensions can be dynamically unstable to gravitational perturbations~\cite{Santos:2015iua, Figueras:2015hkb, Figueras:2017zwa}.
It was first shown for the case of black strings and black $p$-branes~\cite{Gregory:1993vy}.
The nonlinear evolution of the instability of a five-dimensional black string leads to a sequence of black hole bulges connected by black strings that become ever thinner over time.
These thin strings pinch off within a finite time~\cite{Lehner:2010pn}.

The instability for some timelike singularities has been proved.
The linear stability under gravitational perturbations of the negative mass Schwarzschild spacetime was first considered~\cite{Gibbons:2004au}.
Unstable modes were found for not only the negative mass Schwarzschild spacetime, but also the Reissner-Nordstr\"{o}m (RN) and Kerr naked singularities~\cite{Gleiser:2006yz, Dotti:2006gc, Dotti:2008yr}.
This indicates that these timelike naked singularities cannot be the endpoint of gravitational collapse, or even if they can be formed temporarily, they are unstable.
It is essential to investigate what physical phenomena occur in the presence of these timelike naked singularities to conclusively support the weak cosmic censorship conjecture.

Despite these various efforts, the validity of the weak cosmic censorship conjecture remains an open question. Quantum gravity seems to solve this question; however, it is still meaningful to consider what kinds of phenomena need to be considered in order for the weak cosmic censorship conjecture to be effective, even if a naked singularity occurs in the framework of general relativity. Here, we consider the case where a naked singularity exists.

Even if some counterexamples exist, it may be possible that the weak cosmic censorship conjecture would be effectively preserved, if a temporal naked singularity will evolve into a non-singular physical object in a short time.
Note that the occurrence of a naked singularity of limited duration is called ``cosmic flashing''~\cite{Ford:1990ae, Ford:1992ts}.
The following natural question then arises: ``When there exists a naked singularity, does it evolve into a black hole via some physical process?''
We answer this question with a rigorous and detailed analysis in this paper.
One result is following: An extreme RN black hole can be formed at a finite advanced time as a result of the gravitational collapse of a charged thin shell.

An overcharged RN spacetime is also interesting in the context of ultrahigh energy collisions of particles and shells~\cite{Patil:2011uf,Nakao:2013uj}.
In particular, a RN spacetime that is slightly more charged than extremality is capable of accelerating particles and shells so that the centre-of-mass energy of two such colliding objects can be very high and such high-energy reactions can be exposed to observations in the absence of horizons.

In this paper, in order to clarify the possibility of a transition from a naked singularity to a black hole, we analyze a RN spacetime with a charged thin shell, which provides the exact solution of the Einstein equations including the back reaction.
Although we do not show it here, it can be seen from a simple calculation that by dropping charged test particles into a RN naked singularity, it is possible to form an event horizon.
However, to find out what really happens, the inclusion of the self-energy of the test particles or their back reaction is essential.
Hence we shall analyze the gravitational collapse of a charged thin shell (or a singular hypersurface layer), which solves the Einstein equations.
This setting will yield an appropriate analysis.
Note that determining the kind of motion the charged thin shell in this setting takes is not trivial.
Because there is a forbidden region, the charged thin shell does not necessarily collapse into a black hole.

The plan of this paper is as follows.
In Sec.~\ref{sec2}, we present the equations of motion of a charged thin shell.
We present our analysis in Secs.~\ref{sec3}, \ref{sec4}, and \ref{sec5}.
The gravitational collapse of the charged thin shell onto a naked singularity leads to the formation of an event horizon.
This supports the effective validity of the weak cosmic censorship conjecture.
The evolution history of the charged thin shell is shown using the Penrose diagram, which provides the global structure of the present dynamical spacetime; i.e., a black hole is formed by the gravitational collapse of a charged thin shell onto a naked singularity.
We carefully consider the existence of the process, which forms an extreme black hole at a finite advanced time.
We summarize our results and provide some concluding remarks in Sec.~\ref{sec6}.
We use the geometrical units, i.e., $G=c=1$.

\section{Equations of motion}
\label{sec2}
We consider a spherically symmetric electrovac spacetime with an infinitely thin charged shell, described by a timelike hypersurface $\Sigma$, which divides the spacetime $\left( \mathcal{M} , g \right)$ into two regions---the inside and outside.
From the Birkhoff theorem, each region is described by the RN metric:
\begin{eqnarray}
	ds_{\pm}^2 = -f_{\pm} dt_{\pm}^2 + \frac{1}{f_{\pm}} dr^2 + r^2 d \Omega ^2 \, ,
\end{eqnarray}
where
\begin{eqnarray}
	f_{\pm} := 1 - \frac{2 M_{\pm}}{r} + \frac{Q_{\pm} ^2}{r^2} \, ,
\end{eqnarray}
and $d \Omega ^2$ denotes a metric on the unit two-sphere.
Here, we have used the subscript $\pm$ for the variables in the exterior and interior regions, respectively.
Since the circumference radius $r$ is continuous even at the shell surface, we do not apply the subscript $\pm$ to it.

$M$ and $Q$ denote the mass and electric charge of the RN spacetime, respectively.
When $ M \geq | Q | $, the spacetime describes a black hole.
The case $ M = | Q | $ is referred to as an extreme RN black hole, while if $ M < | Q | $ the spacetime describes a naked singularity.

In our setting, since we are discussing the transition from a naked singularity to a black hole, we assume that the masses and electric charges in the interior and exterior geometries satisfy the conditions $ M_{-} < | Q_{-} | $ and $ M_{+} \geq | Q_{+} | $, respectively.

The singular hypersurface $\Sigma$ is described by two-dimensional coordinates $(t_\pm, r)$, with the parametric equations such that $t_\pm =  T_{\pm}\left( \tau \right)$ and $r=R\left( \tau \right)$, where $\tau$ is the proper time of the shell.
The induced metric $h_{ij}$ on $\Sigma$ is given by
\begin{eqnarray}
	ds_{\Sigma}^{2} :=h_{ij}dy^i dy^j
	= -d\tau ^{2} + R^{2} \left( \tau \right) d \Omega \, ,
\end{eqnarray}
with the intrinsic coordinates $y^{i} = \left( \tau , \theta , \phi \right)$ of the hypersurface $\Sigma$.
The four velocities of the thin shell in both the interior and the exterior coordinates are given by 
\begin{eqnarray}
	u_{\pm}^{\mu} \partial _{\mu} = \dot{T}_{\pm} \partial _{t_{\pm}} + \dot{R} \partial _{r} \, ,
\end{eqnarray}
where the overdot denotes differentiation with respect to the proper time $\tau$.
Since the four velocities in both coordinates are normalized as $u_{\pm}^{\mu} u_{\pm \mu} = -1$,
we find
\begin{eqnarray}
	\dot{T}_{\pm} = \frac{\epsilon _{{\rm t \pm}} \sqrt{\dot{R}^{2} + f_{\pm}}}{f_{\pm}} \, ,
	\label{eq005}
\end{eqnarray}
where the signs $\epsilon _{{\rm t}\pm}=1$ or $-1$, which represents the sign of $\dot{T}_{\pm} f_{\pm}$.
We do not assume that $\dot{T}_{+}$ is always positive, since we also consider the interior region of the black hole spacetime ($f_{+} < 0$).
It is important to fix the directions of normal vectors of the thin shell, which have explicit forms that we show below~\cite{Lake:1978zz}.
Note that the time coordinates $t_{\pm}$ are not continuous at the shell radius $r=R$.

The induced metric $h_{ij}$ is continuous at $\Sigma$, i.e.,
\begin{eqnarray}
	\left[ h_{ij} \right]_\pm = 0 \, ,
\end{eqnarray}
where the square brackets ($[~]_\pm$) denote  the difference between the variables evaluated on the both sides of  $\Sigma$, i.e., $[X]_\pm := X_+-X_-$.
This junction condition fixes the relation between the two time coordinates, $t_{-}$ and $t_{+}$, at $\Sigma$.

The unit normal vectors of the thin shell in both coordinates are given by
\begin{eqnarray}
	n_{\pm \mu} dx^{\mu} = \epsilon _{\pm} \left( - \dot{R} dt_{\pm} + \dot{T}_{\pm} dr \right) \, ,
\end{eqnarray}
where the overall signs $\epsilon _{\pm}=1$ or $-1$ show the possible directions of the normals that appear in the equations of motion through calculations for the extrinsic curvature~\cite{Sato:1986}.
This normal vector must satisfy $u_\pm^{\mu}n_{\pm \mu} = 0$ and $n_{\pm}^{\mu} n_{\pm \mu} = 1$.

The stress energy tensor $S_{ij}$ of the thin shell determines the jump of the extrinsic curvature $K_{ij}$ of $\Sigma$ as
\begin{eqnarray}
	\left[ K_{ij} \right]_\pm - \left[ K \right]_\pm h_{ij} = -8\pi S_{ij}\, ,
	\label{eq010}
\end{eqnarray}
where $K=h^{ij}K_{ij}$ is the trace of the extrinsic curvature.
From the difference of the Hamiltonian constraint equations on both sides of $\Sigma$ with the junction condition (\ref{eq010}), we find
\begin{eqnarray}
	S_{ij} \{K^{ij}\}_\pm = \left[ T_{\mu \nu} n^{\mu} n^{\nu} \right]_\pm \, ,
	\label{eq020}
\end{eqnarray}
where $T^{\mu \nu}$ is the energy-momentum tensor associated with spacetime.
The brackets $\{\}_\pm$ denote the mean value of the variables evaluated on both sides of $\Sigma$, i.e., $\{X\}_\pm := \left( X_+ +X_- \right) /2$.
Using the momentum constraint, we find the following conservation equation of the shell energy-momentum:
\begin{eqnarray}
	D_{i} S^{i}_{j} = - \left[ T_{\mu \nu} e_{j}^{\mu} n^{\nu} \right]_\pm \, ,
	\label{eq030}
\end{eqnarray}
where $e_{i}^{\mu}$ is the set of associated tangent basis vectors of the intrinsic coordinates $y^{i}$, and $D_{i}$ is a covariant derivative on $\Sigma$.

In what follows, we assume that the charged thin shell is composed of a pressureless matter fluid, i.e., 
\begin{eqnarray}
	S^{ij} = \sigma u^{i} u^{j} \, ,
\end{eqnarray}
where $\sigma$ and $u^{i}$ are  the surface mass density and the velocity field of the charged thin shell.
Each component of the  junction condition~(\ref{eq010}) then reads as follows:
\begin{eqnarray}
	&& \left[ \frac{\epsilon \epsilon _{\rm t}}{\sqrt{\dot{R}^{2} + f}} \left( \ddot{R} + \frac{1}{2}\frac{df}{dr} \right) + \frac{\epsilon \epsilon _{\rm t} \sqrt{\dot{R}^{2} + f}}{R} \right]_\pm = 0 \, ,
	\label{eq040} \\
	&& \left[ - \frac{\epsilon \epsilon _{\rm t} \sqrt{\dot{R}^{2} + f}}{R} \right]_\pm = 4 \pi \sigma \, .
	\label{eq050}
\end{eqnarray}
The energy-momentum tensor $T^{\mu \nu}$ associated with the RN spacetime is 
\begin{eqnarray}
	- T^{t}_{\ t} = - T^{r}_{\ r} = T^{\theta}_{\ \theta} = T^{\phi}_{\ \phi} = \frac{Q^{2}}{8 \pi r^{4}} \, ,
\end{eqnarray}
and the other components vanish.
Relation~(\ref{eq020}) gives
\begin{eqnarray}
	\left\{\frac{\epsilon \epsilon _{\rm t}}{\sqrt{\dot{R}^{2} + f}} \left( \ddot{R} + \frac{1}{2}\frac{df}{dr} \right) 
	\right\}_\pm = 0 \, .
	\label{eq060}
\end{eqnarray}
For the comoving coordinates employed in the induced metric ($u^{i} \partial _{i} = \partial _{\tau}$), Eq.~(\ref{eq030}) is integrated as follows:
\begin{eqnarray}
	m := 4\pi R^2 \sigma = {\rm constant}\, ,
	\label{eq070}
\end{eqnarray}
where $m$ is the conserved rest mass of the charged thin shell.

From the basic equations~(\ref{eq040}), (\ref{eq050}), (\ref{eq060}), and (\ref{eq070}),  we find the equation of motion as
\begin{eqnarray}
	\sqrt{\dot{R}^{2} + f_{+}} = \epsilon _{+} \epsilon _{{\rm t}+} F(R) \, ,
	\label{eq080}
\end{eqnarray}
where
\begin{eqnarray}
	F(R) := \frac{R}{2m} \left( f_{-} - f_{+} \right) - \frac{m}{2R} \, ,
	\label{eq090}
\end{eqnarray}
and the possible range of the radius of the charged thin shell is restricted as
\begin{eqnarray}
	\epsilon _{+} \epsilon _{{\rm t}+} F(R) \geq  0 \, ,
	\label{eq100}
\end{eqnarray}
and
\begin{eqnarray}
	\epsilon _{-} \epsilon _{{\rm t}-} \left( F(R) + \frac{m}{R} \right) \geq 0 \, .
	\label{eq105}
\end{eqnarray}
Furthermore, the radius of the charged thin shell is restricted as
\begin{eqnarray}
	F^2 (R)- f_{+}(R) \geq 0 \, .
	\label{eq110}
\end{eqnarray}
We comprehensively derived equations of motion~(\ref{eq005}) and (\ref{eq080}), leaving the choice of the directions of the normal vectors.
The conditions~(\ref{eq100}) and (\ref{eq105}) of the equations of motion are shown clearly.

The energy and electric charge of the charged thin shell are denoted by $E$ and $q$, respectively.
When we discuss the gravitational collapse of the charged thin shell, the equations of motion will be described by the following physical parameters: $M_{-}$, $Q_{-}$, $E$, $q$, and $m$.
The energy and electric charge of the charged thin shell have relations with the masses and electric charges of the interior and exterior metrics as
\begin{eqnarray}
	E &=& M_{+} - M_{-} \, , \\
	q &=& Q_{+} - Q_{-} \, .
	\label{eq120}
\end{eqnarray}

To simplify the problem, we assume that $\dot{T}_{\pm} f_{\pm}$ are positive.
The validity of these assumptions will be explained at the end of Sec.~\ref{sec2}.
Thus, the signs are set to be positive $\epsilon _{{\rm t}\pm} = 1$.
We further assume that the overall signs of the normal vectors of the thin shell are positive $\epsilon _{\pm} = 1$.
Under the setting $\epsilon _{{\rm t}\pm} = 1$, the assumptions $\epsilon _{\pm} = 1$ indicate that the normal vectors point to a direction in which the radial coordinate increases.
We assume that the rest mass $m$ of the charged thin shell is positive.
Then, Eqs.~(\ref{eq100}) and (\ref{eq105}) reduce to the single restriction
\begin{eqnarray}
	F(R) \geq 0 \, .
	\label{eq125}
\end{eqnarray}

At this juncture, we shall summarize our assumptions about the parameter region.
First, we assume that the energy $E$ of the charged thin shell is positive.
We also assume that the electric charge $Q_{-}$ of the interior metric is positive without loss of generality.
In order to further simplify our problem, we restrict the rest mass $m$ and the electric charge $q$ of the charged thin shell to the following parameter region, for a given mass $M_{-}$ and electric charge $Q_{-}$ of the interior spacetime.
These assumptions are written as 
\begin{eqnarray}
	0 < m \leq Q_{-} \, ,
	\label{eq130}
\end{eqnarray}
and
\begin{eqnarray}
	-Q_{-} - \sqrt{Q_{-}^2 - m^2} \leq q \leq -Q_{-} + \sqrt{Q_{-}^2 - m^2} \, .
	\label{eq140}
\end{eqnarray}
Under these assumptions, the conditions do not forbid the charged thin shell from reaching the singularity.
We can see in this paper that Eqs.~(\ref{eq130}) and (\ref{eq140}) are sufficient conditions for a charged thin shell to collapse into a RN black hole.

Equation~(\ref{eq080}) is rewritten in the form of one-dimensional particle motion~\cite{Cruz:1967,Kuchar:1968,Chase:1970}:
\begin{eqnarray}
	\dot R^2 + \mathcal{V} = 0\,,
	\label{eq145}
\end{eqnarray}
where we have defined the ``effective'' potential
\begin{eqnarray}
	\mathcal{V} := f_+(R) - F^2(R)\,.
	\label{eq146}
\end{eqnarray}
Since Eqs.~(\ref{eq130}) and (\ref{eq140}) are satisfied, Eqs.~(\ref{eq080}) and (\ref{eq145}), both equations of motion, are equivalent.
By virtue of Eqs.~(\ref{eq130}) and (\ref{eq140}), the function $F$ becomes positive for all $R > 0$.
From our ansatz $M_{+} \geq | Q_{+} |$, we have the following condition as well:
\begin{eqnarray}
	| Q_{-} + q | \leq M_{-} + E \, .
	\label{eq150}
\end{eqnarray}

Substituting Eq.~(\ref{eq080}) into Eq.~(\ref{eq005}) yields the following equation:
\begin{eqnarray}
	\dot{T}_{+} f_{+} = \epsilon _{+} F \, .
	\label{eq155}
\end{eqnarray}
If Eqs.~(\ref{eq130}) and (\ref{eq140}) are satisfied, $\dot{T}_{+} f_{+} \neq 0$.
In such parameter regions, the sign of $\dot{T}_{+}$ in the region $f_{+} < 0$ does not change.
Therefore it is possible to assume $\dot{T}_{+}f_{+} > 0$ in the region $f_{+} < 0$.
Also, since $f_{-}$ is always positive, the hypersurfaces $\left\{ t_{-} = {\rm constant} \right\}$ are spacelike.
In addition, it is possible to assume that $\dot{T}_{-}$ is positive.
As a consequence, $\dot{T}_{-} f_{-} > 0$ holds.

\section{Transition from  a naked singularity to a black hole}
\label{sec3}
\begin{figure}[tb]
		\begin{tabular}{ cc }
			\includegraphics[width=4.3cm]{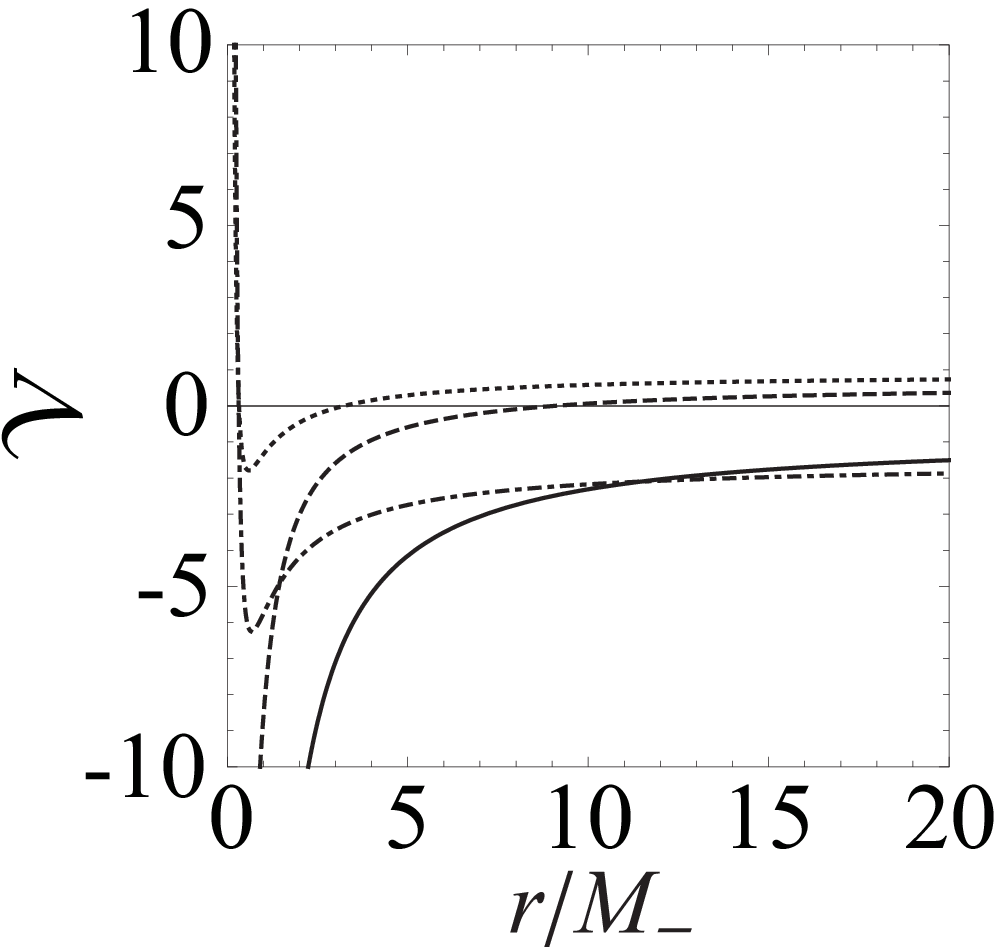}
		\end{tabular}
	\caption{\footnotesize{The effective potential $\mathcal{V}$ for radial motion in the spacetime for $Q_{-} = 2M_{-}$, $E = 0.4M_{-}$, $q = -0.8M_{-}$, $m = 0.3M_{-}$ (solida line), $E = 0.3M_{-}$, $q = -0.8M_{-}$, $m = 0.5M_{-}$ (dashed line), $E = 0.8M_{-}$, $q = -0.3M_{-}$, $m = 0.5M_{-}$ (dash-dotted line), and $E = 0.5M_{-}$, $q = -M_{-}$, $m = 1.5M_{-}$ (dotted line). In all the four cases physical quantities satisfy $0 < | Q_{-} + q | < M_{-} + E$.}}
		\label{fig1}
\end{figure}
\begin{figure}[tb]
		\begin{tabular}{ cc }
			\includegraphics[width=4.3cm]{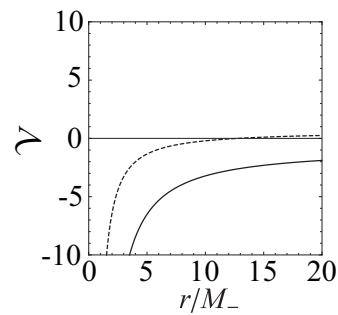}
		\end{tabular}
	\caption{\footnotesize{The effective potential $\mathcal{V}$ for radial motion in the spacetime for $Q_{-} = 2M_{-}$, $E=0.4M_{-}$, $q=-2M_{-}$, $m=0.3M_{-}$ (solid line) and $E=0.3M_{-}$, $q=-2M_{-}$, $m=0.5M_{-}$ (dashed line). In the both cases physical quantities satisfy $0 = | Q_{-} + q | < M_{-} + E$.}}
		\label{fig2}
\end{figure}
\begin{figure}[tb]
		\begin{tabular}{ cc }
			\includegraphics[width=4.3cm]{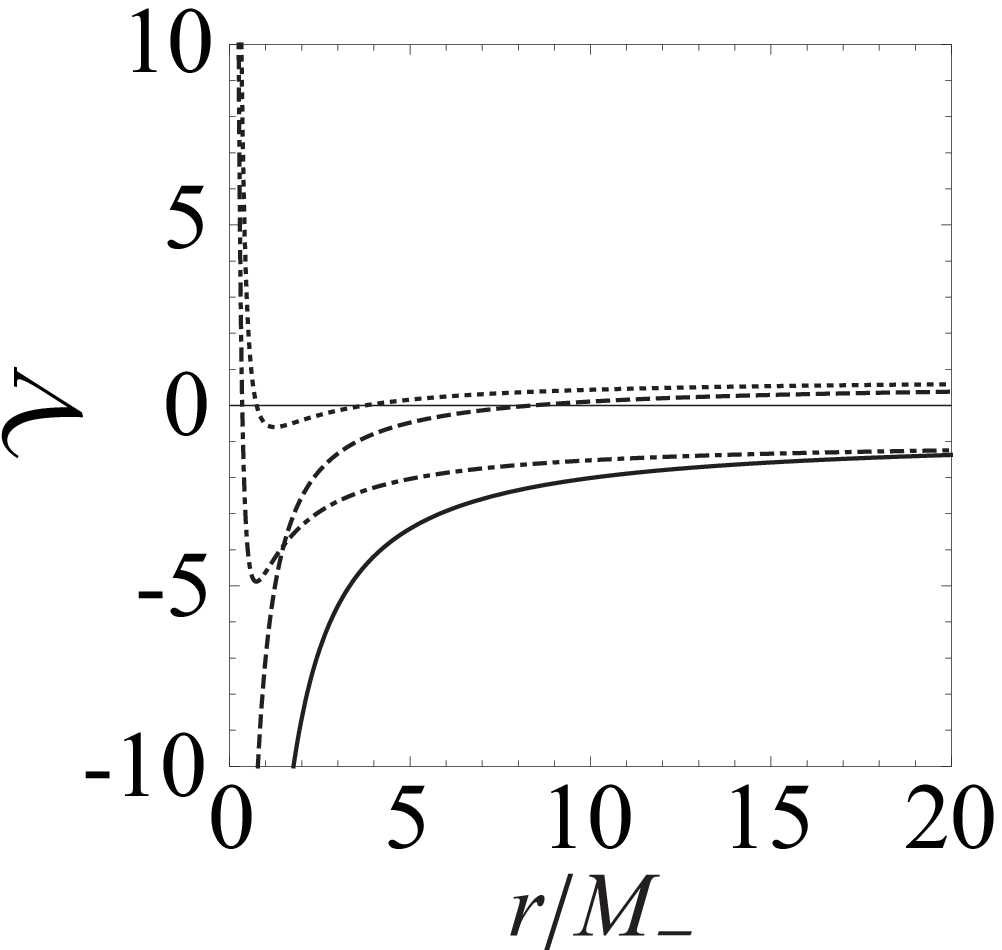}
		\end{tabular}
	\caption{\footnotesize{The effective potential $\mathcal{V}$ for radial motion in the spacetime for $Q_{-} = 2M_{-}$, $E=0.4M_{-}$, $q=-0.6M_{-}$, $m=0.3M_{-}$ (solid line), $E=0.3M_{-}$, $q=-0.7M_{-}$, $m=0.5M_{-}$ (dashed line), $E=0.7M_{-}$, $q=-0.3M_{-}$, $m=0.5M_{-}$ (dash-dotted line), and $E=0.5M_{-}$, $q=-0.5M_{-}$, $m=M_{-}$ (dotted line). In all the four cases physical quantities satisfy $0 < | Q_{-} + q | = M_{-} + E$.}}
		\label{fig3}
\end{figure}
\begin{figure}[tb]
		\begin{tabular}{ cc }
			\includegraphics[width=5.5cm]{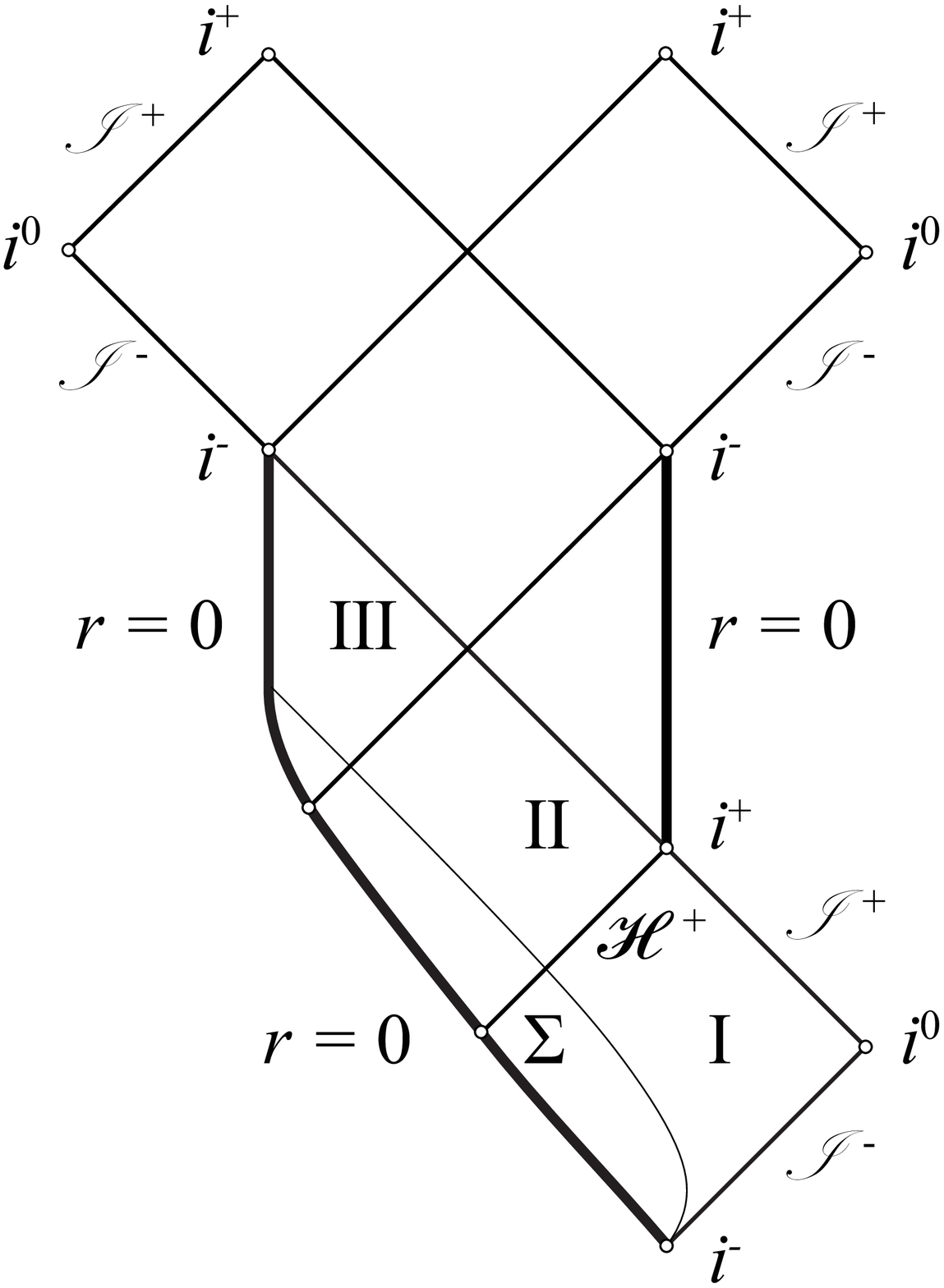}
		\end{tabular}
	\caption{\footnotesize{Penrose diagram of the spacetime for $0 < | Q_{-} + q | < M_{-} + E$ and $m \leq \min \{ |q|,E \}$. The thin curve is the crash trajectory of the thin shell.}}
		\label{fig4}
\end{figure}
\begin{figure}[tb]
		\begin{tabular}{ cc }
			\includegraphics[width=5.5cm]{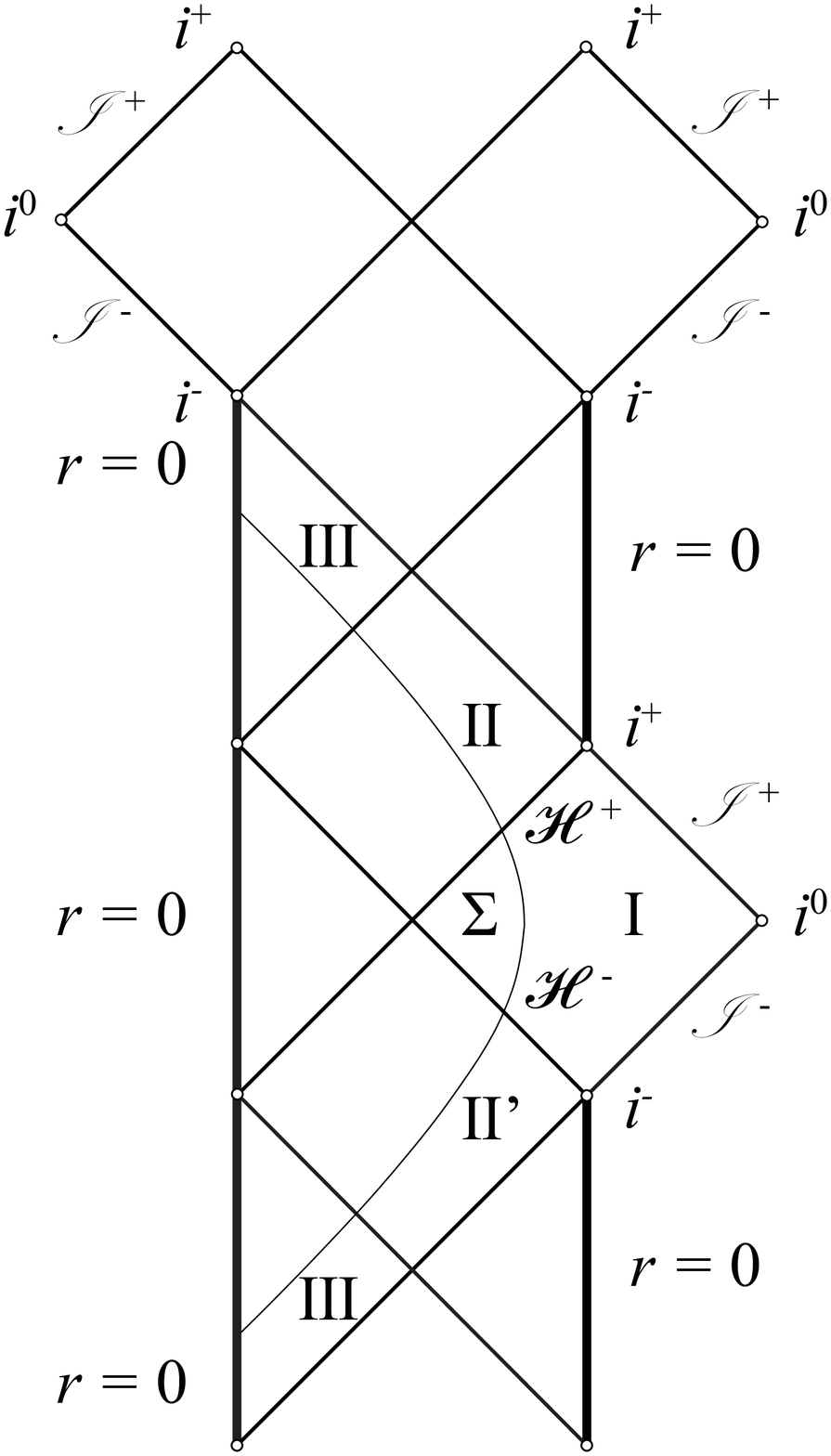}
		\end{tabular}
	\caption{\footnotesize{Penrose diagram of the spacetime for $0 < | Q_{-} + q | < M_{-} + E$ and $E < m \leq |q|$. The thin curve is the clap trajectory of the thin shell.}}
	\label{fig5}
\end{figure}
\begin{figure}[tb]
		\begin{tabular}{ cc }
			\includegraphics[width=5.1cm]{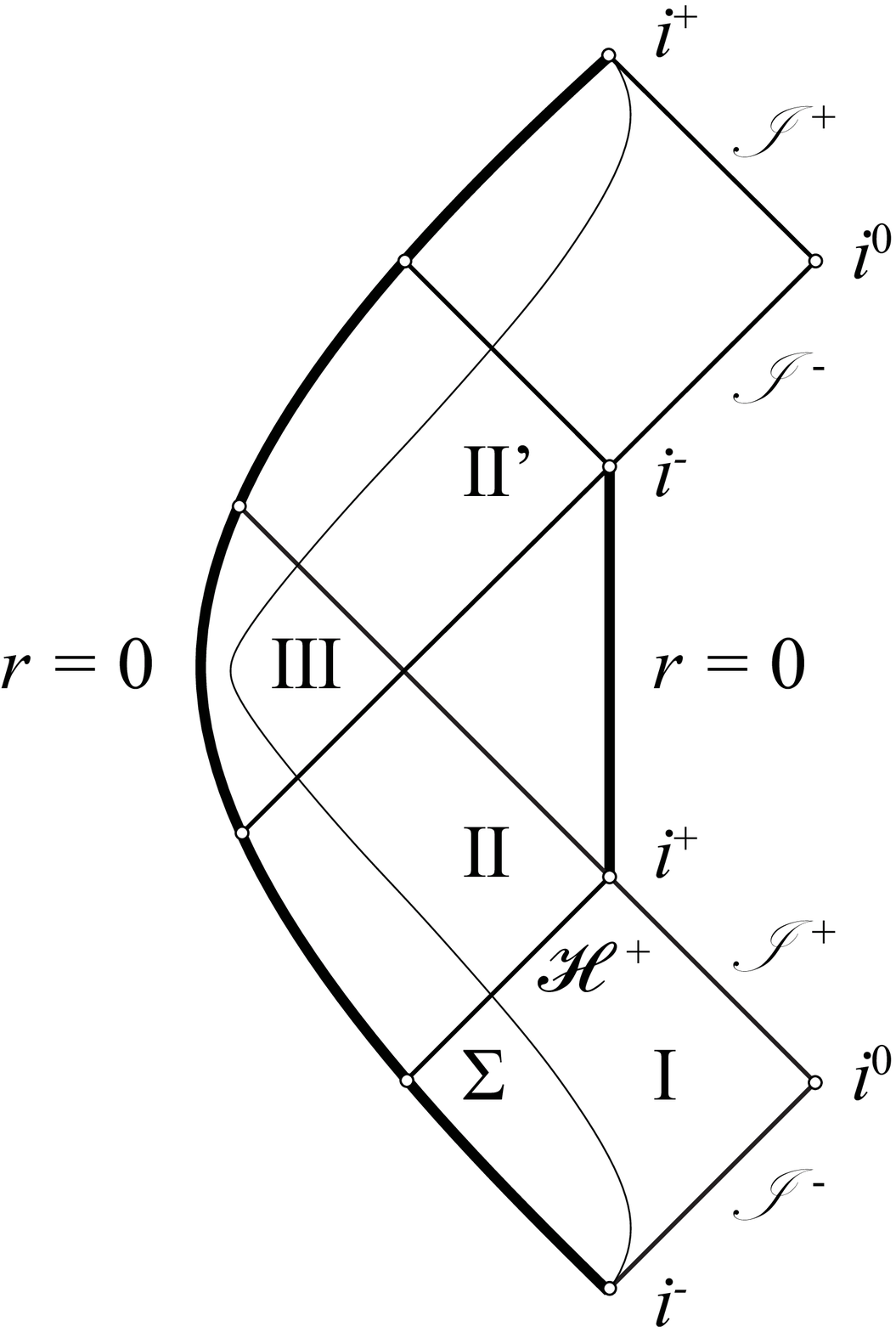}
		\end{tabular}
	\caption{\footnotesize{Penrose diagram of the spacetime for $0 < | Q_{-} + q | < M_{-} + E$ and $|q| < m \leq E$. The thin curve is the flyby trajectory of the thin shell.}}
	\label{fig6}
\end{figure}
\begin{figure}[tb]
		\begin{tabular}{ cc }
			\includegraphics[width=5.0cm]{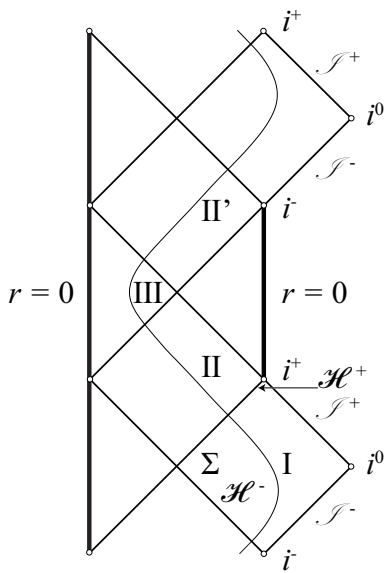}
		\end{tabular}
	\caption{\footnotesize{Penrose diagram of the spacetime for $0 < | Q_{-} + q | < M_{-} + E$ and $\max \{ |q|,E \} < m$. The thin curve is the bound trajectory of the thin shell.}}
	\label{fig7}
\end{figure}
We next consider the the formation of a RN black hole via the gravitational collapse of a charged thin shell onto a RN naked singularity.
We investigate the system described by Eqs.~(\ref{eq005}) and (\ref{eq080}).
First, we analyze Eq.~(\ref{eq080}) to understand the radial motion of the thin shell with respect to proper time.
$T_{\pm}\left( \tau \right)$ can be derived from Eq.~(\ref{eq005}).
To clarify the existence of the solution of the system of the equations of motion, we rewrite them as a normal system~\cite{pontryagin1962ordinary} of ordinary differential equations.
After clarifying the solution's existence, we visually confirm the characteristics of the motion by illustrating the effective potential $\mathcal{V}$.

We then describe the initial value problem as
\begin{eqnarray}
	\dot R \left( \tau \right) &=& V \left( R \left( \tau \right) \right) \, ,
	\label{eq165}
\end{eqnarray}
and
\begin{eqnarray}
	R \left( \tau _{0} \right) &=& r_{0} \, ,
	\label{eq170}
\end{eqnarray}
where
\begin{eqnarray}
	V := \epsilon _{\rm r} \sqrt{F^2 - f_{+}} \, ,
	\label{eq180}
\end{eqnarray}
with $\tau_0$ and $r_{0}$ as the initial values.
The function $V$ is defined on the domain, which is determined by Eq.~(\ref{eq110}).
The sign $\epsilon _{\rm r}$ changes when the shell reaches a $r$-turning point.
Since the function $V$ does not explicitly depend on the proper time $\tau$, this is a normal autonomous system.
While analyzing motion as a potential problem, the sign is commonly taken into account implicitly.
Let $I$ be an open interval, s.t.,
\begin{eqnarray}
	I = \left\{r \mid \left( F^2 - f_{+} \right) \left( r \right) > 0 \right\} \, .
	\label{eq190}
\end{eqnarray}
We can see that $I$ is simply connected.
For the initial condition given by Eq.~(\ref{eq170}), there is a unique maximally extended solution $R(\tau)$ in the range $I$.
Choosing $\epsilon _{\rm r} = -1$, $R(\tau)$ is a strictly monotonically decreasing function in the domain $(\tau_{1},\tau_{2})$, where we can see
\begin{eqnarray}
	\lim_{\tau \downarrow \tau_{1}} R\left( \tau \right) &=& \sup I \, , \\
  	\label{eq200}
	\lim_{\tau \uparrow \tau_{2}} R\left( \tau \right) &=& \inf I \, .
  	\label{eq210}
\end{eqnarray}
Note that $\tau _{1}$ may be $-\infty$ and $\tau _{2}$ may be $+\infty$.

The existence of the zeros $r_{\rm e}$ and $r_{\rm c}$ $\left( r_{\rm e} \geq r_{\rm c} \right)$ of $f_{+}$ is guaranteed by Eq.~(\ref{eq150}).
If the thin shell can reach this radius $r_{\rm e}$ and go beyond, a RN black hole is formed.
This is confirmed later by analyzing the global structure of the spacetime.
By a simple calculation we find that $\left( F^2 - f_{+} \right) \left( r_{\rm e} \right) > 0$ and $\left( F^2 - f_{+} \right) \left( r_{\rm c} \right) > 0$.
This means that $\left\{ r_{\rm e}, r_{\rm c} \right\} \subset I$.

The condition that there exists a zero $r_{\rm{sup}}$ of the function $V$, which is larger than $r_{\rm e}$, is equivalent to the condition that $\left( F^2 - f_{+} \right)$ in the limit of infinity is negative.
In fact, the graph of $\left( F^2 - f_{+} \right)$ can have only one extremal point, and it is concave if there is an extreme.
This condition can be written as follows:
\begin{eqnarray}
	m > E \, .
  	\label{eq215}
\end{eqnarray}
Note that we have used
\begin{eqnarray}
	\lim_{r \to \infty} \mathcal{V} = \frac{\left( m + E \right) \left( m - E \right)}{m^2} \, .
  	\label{eq217}
\end{eqnarray}
From these logics, we will take the initial value $r_{0}$ in the following range:
\begin{eqnarray}
	r_{\rm e} < r_{0} < 
	\begin{cases}
	r_{\rm{sup}} & {\rm if }~ m > E \, ,  \\
	\infty &  {\rm if }~  m \leq E \, . 
	\end{cases}
  	\label{eq220}
\end{eqnarray}
Then the orbit $\mathcal{O}$ of the maximal set of solutions of the initial value problem must include $\left\{ r_{\rm e}, r_{\rm c} \right\}$,
\begin{eqnarray}
	\left\{ r_{\rm e}, r_{\rm c} \right\} \subset \mathcal{O} \left( r_{0} \right) := \left\{ R \left( \tau \right) \mid \tau \in \left( \tau _{1} , \tau _{2} \right) \right\} \, .
	\label{eq230}
\end{eqnarray}
For the initial radius $r_{0}$ to be arbitrarily large, we find from Eq.~(\ref{eq220}) that the rest mass must be less than or equal to the energy, $m \leq E$.

The condition that there exists a zero $r_{\rm inf}$ of the function $V$, which is smaller than $r_{\rm c}$, is equivalent to the condition that $\left( F^2 - f_{+} \right)$ in the limit of $r = 0$ is negative.
This can be written as follows:
\begin{eqnarray}
	m > |q| \, .
  	\label{eq234}
\end{eqnarray}
Note that we have used
\begin{eqnarray}
	\lim_{r \downarrow 0} \mathcal{V} = 
	\begin{cases}
	\frac{\left( m + q \right) \left( m - q \right)}{\left| m + q \right| \left| m - q \right|} \cdot \left( + \infty \right) & {\rm if }~ m \neq |q| \, ,  \\
	- \infty &  {\rm if }~  m = |q| \, . 
	\end{cases}
  	\label{eq235}
\end{eqnarray}
We have been able to confirm the existence of the solution strictly by the discussion so far.

In Sec.~\ref{sec2}, the equations of motion of the thin shell were described in spherical coordinates.
As is widely known, a RN black hole can be maximally extended.
A RN naked singularity is already inextendible in spherical coordinates.
Trajectories of the thin shell will be analyzed in terms of ingoing and outgoing Eddington-Finkelstein coordinates.
In order to make it easier to confirm the formation of an event horizon, we analyze the gravitational collapse of the thin shell using the coordinate systems of the exterior region.

The global structure of the spacetime is classified based on the trajectory of the thin shell.
We focus on the existence of the $r$-turning points that determine whether a trajectory reaches the singularity or infinity.
From Eqs.~(\ref{eq215}) and (\ref{eq234}), the thin shell can have four types of trajectories:
\begin{itemize}
  \item Crash if $m \leq \min \{ |q|, E \}$\,,
  \item Clap if $E < m \leq |q|$\,,
  \item Flyby if $|q| < m \leq E$\,,
  \item Bound if $\max \{ |q|, E \} < m$\,.
\end{itemize}
Here, the effective potential $\mathcal{V}$ for radial motion is illustrated in Figs.~\ref{fig1}--\ref{fig3} to visually deepen an understanding of the characteristics of motion.
The radial motion of the crash, clap, flyby, and bound types correspond to the effective potential shown, respectively, by the solid, dashed, dash-dotted, and dotted lines in Figs.~\ref{fig1}--\ref{fig3}.

From the viewpoint of the law of conservation of energy, we give an interpretation of the conditions that determine the trajectory type.
Equation~(\ref{eq080}) expresses the law of conservation of energy~\cite{Kuchar:1968}; this is clear, if we put it in the following form:
\begin{eqnarray}
	m \sqrt{\dot{R}^{2} + f_{-}} - \frac{|q|Q_{-}}{R} + \frac{|q|^2}{2R} - \frac{m^2}{2R} = E \, .
  	\label{eq236}
\end{eqnarray}
The first term in the left-hand side represents the inertial mass of the thin shell.
The second and third terms represent not only the electromagnetic interaction with the RN naked singularity, but also the electromagnetic self-repulsion of the dust composing the thin shell.
The fourth term represents the gravitational interaction of the dust.

We consider condition (\ref{eq215}) under which the thin shell can reach infinity.
The terms of electromagnetic and gravitational interactions are negligible in the law of conservation of energy when the radius of the shell is large enough.
Whether the law of conservation of energy holds depends on the relation between rest mass and energy.

Next, we consider condition (\ref{eq234}) under which the shell cannot reach the singularity.
Intuitively, we expect that if the rest mass is large enough, the contribution of gravitational interaction energy will be immense, and the thin shell will reach the singularity.
However, for the effective potential to diverge positively near the singularity, the condition must be satisfied.
Because the Lorentz factor increases larger near the singularity, the condition differs from that assumed intuitively.

Before further explaining the details of these types, we shall define some terminology.
A black hole region of a spacetime is the set of all events that do not belong to the causal past of the future null infinity: $B := \mathcal{M} \setminus J^{-} \left( \mathscr{I}^{+} \right)$.
The future event horizon is defined as the boundary of the black hole region: $\mathscr{H}^{+} := \partial B$.
In the same sense, the past event horizon is defined as follows: $\mathscr{H}^{-} := \partial W$ where $W := \mathcal{M} \setminus J^{+} \left( \mathscr{I}^{-} \right) $.

\section{Subextreme black hole: $|Q_{+}|<M_{+}$}
\label{sec4}
\begin{figure}[tb]
		\begin{tabular}{ cc }
			\includegraphics[width=3.9cm]{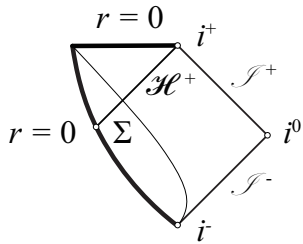}
		\end{tabular}
	\caption{\footnotesize{Penrose diagram of the spacetime for $q = -Q_{-}$ and $m \leq E$. The thin curve is the crash trajectory of the thin shell.}}
	\label{fig8}
\end{figure}
\begin{figure}[tb]
		\begin{tabular}{ cc }
			\includegraphics[width=4.0cm]{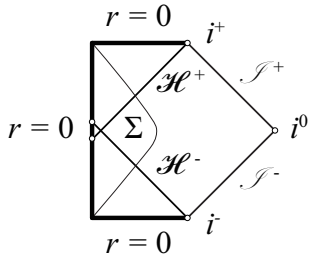}
		\end{tabular}
	\caption{\footnotesize{Penrose diagram of the spacetime for $q = -Q_{-}$ and $E < m$. The thin curve is the clap trajectory of the thin shell.}}
	\label{fig9}
\end{figure}
Here, we consider the case $| Q_{-} + q | < M_{-} + E$.
In particular, for $Q_{-} + q > 0$, we clarify whether a charged thin shell can undergo gravitational collapse onto an overcharged interior object.
We do not limit our discussion solely to a weakly charged thin shell or a slightly overcharged interior object.
The case $Q_{-} + q = 0$ is described at the end of this section.

A crash trajectory of the thin shell is shown in Fig.~\ref{fig4}.
The thin shell collapses from region I of the ingoing Eddington-Finkelstein patch, crosses the horizons, enters region ${\rm I\hspace{-1pt}I\hspace{-1pt}I}$, and hits the curvature singularity $r=0$.
Such trajectories are realized for $m \leq \min \{ |q|, E \}$.
The shell crosses the horizon $\mathscr{H}^{+}$ in an ingoing Eddington-Finkelstein patch and enters region ${\rm I\hspace{-1pt}I}$ where $\dot{R} < 0$.
The shell leaves region ${\rm I\hspace{-1pt}I}$ through the horizon $\mathscr{H}^{-}$ entering region ${\rm I\hspace{-1pt}I\hspace{-1pt}I}$ in the ingoing Eddington-Finkelstein patch.
The trajectory has no $r$-turning point.
Thus the thin shell hits the curvature singularity $r=0$.
Note that the trajectory does not intersect the bifurcation two-sphere.

A clap trajectory of the thin shell is shown in Fig.~\ref{fig5}.
The thin shell expands from region ${\rm I\hspace{-1pt}I\hspace{-1pt}I}$ of an outgoing Eddington-Finkelstein patch.
Such trajectories are realized for $E < m \leq |q|$.
In order to study with this trajectory, it is also necessary to consider Eqs.~(\ref{eq165}), (\ref{eq170}), and (\ref{eq180}) in the case where $\epsilon _{\rm r}$ is plus.
The major difference between the clap and crash trajectories is the existence of the $r$-turning point in region I.

A flyby trajectory of the thin shell is shown in Fig.~\ref{fig6}.
Such trajectories are realized for $|q| < m \leq E$.
The shell crosses the horizon $\mathscr{H}^{+}$ in an ingoing Eddington-Finkelstein patch and enters region ${\rm I\hspace{-1pt}I}$ where $\dot{R} < 0$ continues to hold.
The shell leaves region ${\rm I\hspace{-1pt}I}$ through the horizon $\mathscr{H}^{-}$ entering region ${\rm I\hspace{-1pt}I\hspace{-1pt}I}$ in an ingoing Eddington-Finkelstein patch.
The shell $r$-turns in region ${\rm I\hspace{-1pt}I\hspace{-1pt}I}$.
Thus the shell leaves region ${\rm I\hspace{-1pt}I\hspace{-1pt}I}$ through $\mathscr{H}^{-}$, entering region ${\rm I\hspace{-1pt}I}$' of an outgoing Eddington-Finkelstein patch.
The shell crosses $\mathscr{H}^{+}$ into region I and goes on out to $r = \infty$.

A bound trajectory of the thin shell is shown in Fig.~\ref{fig7}.
The thin shell passes through regions I, ${\rm I\hspace{-1pt}I}$, ${\rm I\hspace{-1pt}I\hspace{-1pt}I}$, and ${\rm I\hspace{-1pt}I}$', but does not hit the curvature singularity $r=0$.
This property of the trajectory is sometimes called oscillation~\cite{Gao:2008jy}.
Such trajectories are realized for $\max \{ |q|, E \} < m$.
The shell crosses the horizon $\mathscr{H}^{+}$, traverses region I\hspace{-1pt}I, and leaves there through $r=r_{\rm c}$.
The shell enters region ${\rm I\hspace{-1pt}I\hspace{-1pt}I}$.
Then the trajectory has its $r$-turn at $r = r_{\rm{inf}}$ and enters region ${\rm I\hspace{-1pt}I}$' of the outgoing patch containing region ${\rm I\hspace{-1pt}I\hspace{-1pt}I}$.
Trips from region I to the next region I are duplicated in the future and in the past.

The analysis so far clarifies the global structure of the spacetimes. We can say that a charged thin shell collapses and an event horizon $\mathscr{H}^{+}$ forms in our system. Thus, there is a dynamical transition from a RN naked singularity to a RN black hole.
We can make this transition without any difficulty.

Although the electric charge of the thin shell must be fine-tuned, $q = -Q_{-}$, even a Schwarzschild black hole can be formed as a result of the transition, as shown in Figs.~\ref{fig8} and \ref{fig9}.
Gravitational collapse of the thin shell occurs, as clarified in the previous discussion.
In this case, the trajectories of the thin shell are of either the crash or the clap type.

\section{Extreme black hole: $|Q_{+}|=M_{+}$}
\label{sec5}
\begin{figure}[tb]
		\begin{tabular}{ cc }
			\includegraphics[width=3.9cm]{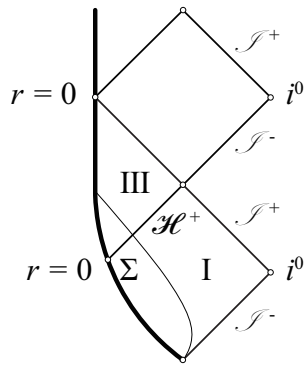}
		\end{tabular}
	\caption{\footnotesize{Penrose diagram of the spacetime for $0 < | Q_{-} + q | = M_{-} + E$ and $m \leq \min \{ |q|,E \}$. The thin curve is the crash trajectory of the thin shell.}}
	\label{fig10}
\end{figure}
\begin{figure}[tb]
		\begin{tabular}{ cc }
			\includegraphics[width=3.9cm]{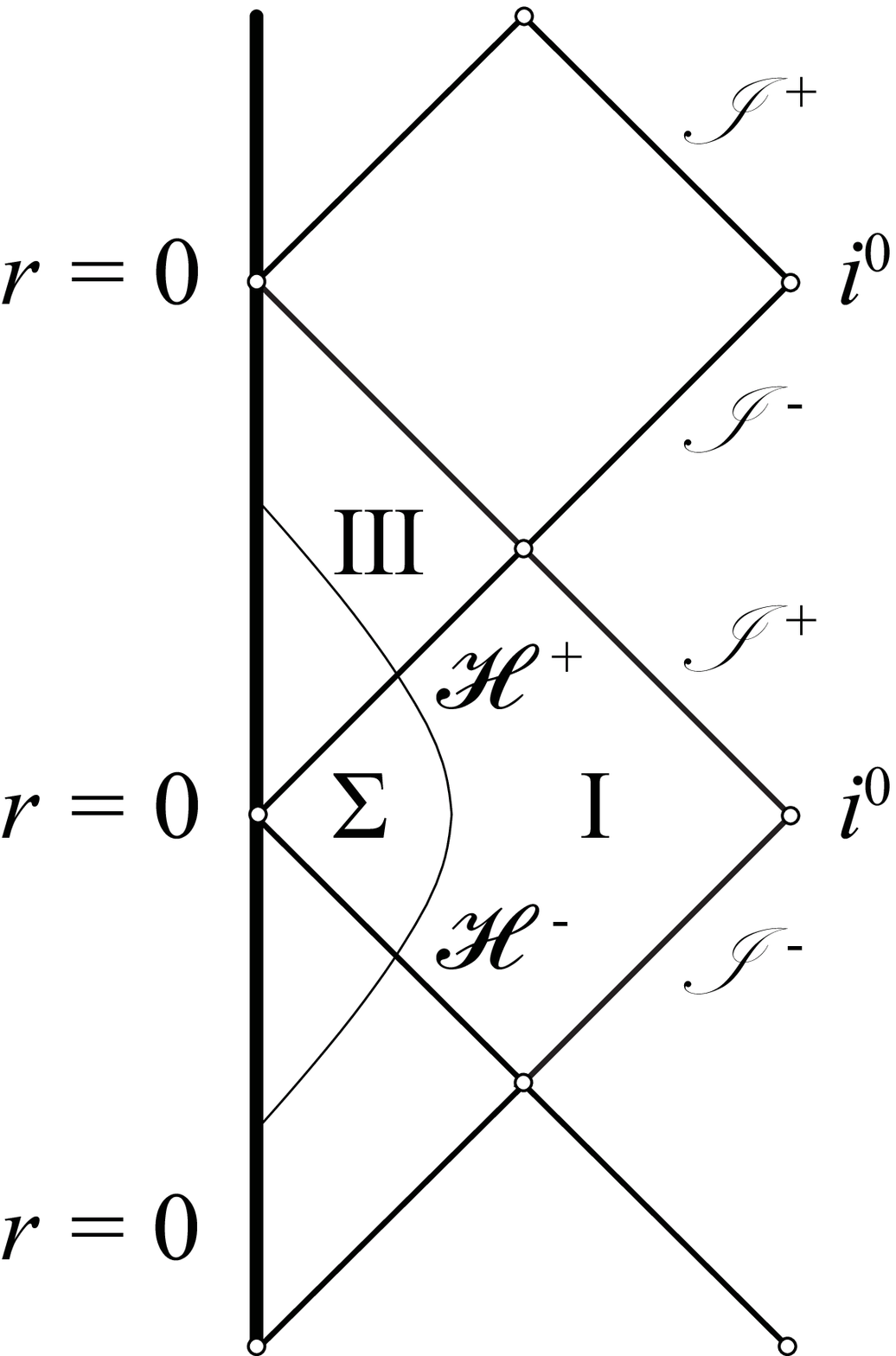}
		\end{tabular}
	\caption{\footnotesize{Penrose diagram of the spacetime for $0 < | Q_{-} + q | = M_{-} + E$ and $E < m \leq |q|$. The thin curve is the clap trajectory of the thin shell.}}
	\label{fig11}
\end{figure}
\begin{figure}[tb]
		\begin{tabular}{ cc }
			\includegraphics[width=3.9cm]{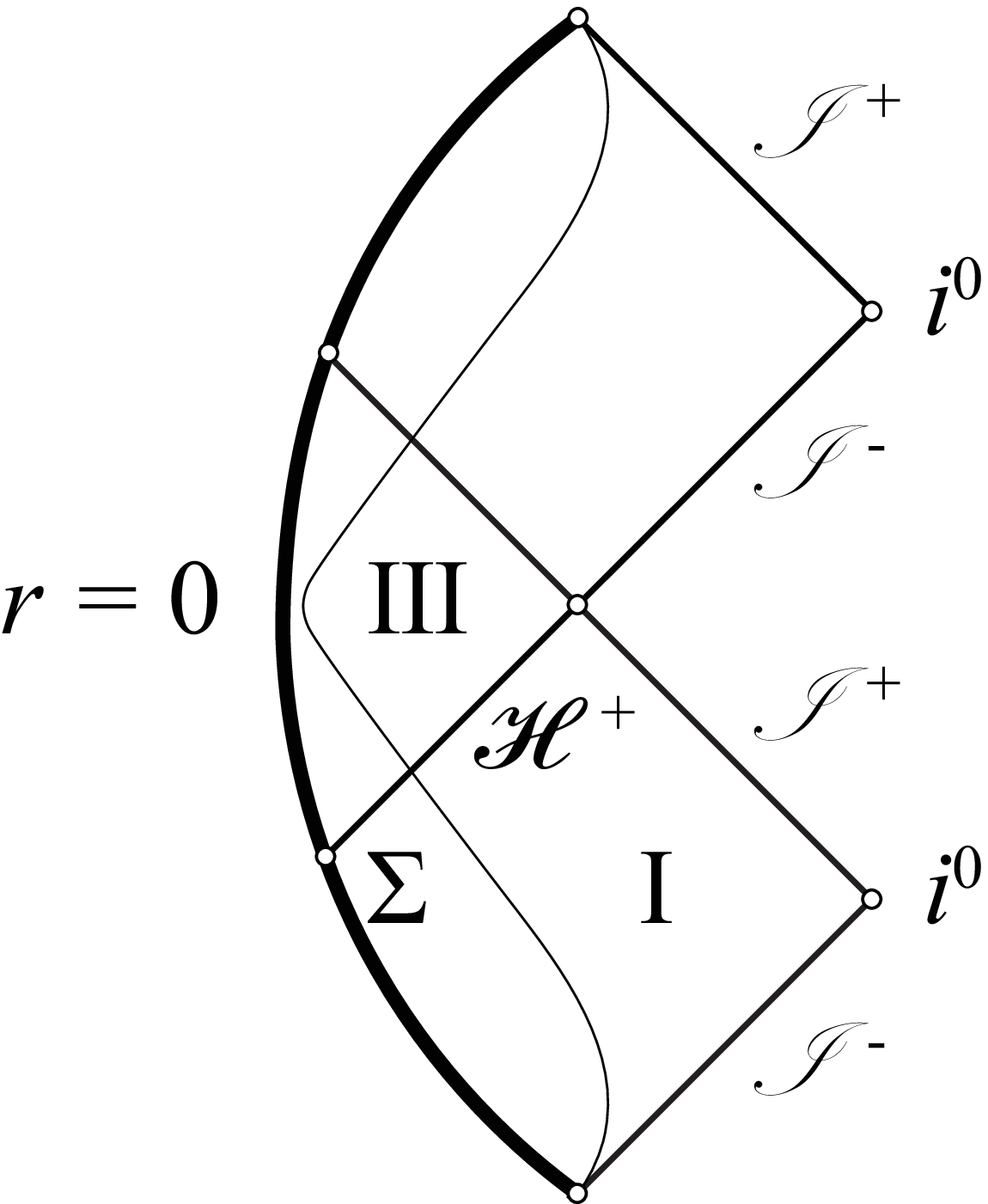}
		\end{tabular}
	\caption{\footnotesize{Penrose diagram of the spacetime for $0 < | Q_{-} + q | = M_{-} + E$ and $|q| < m \leq E$. The thin curve is the flyby trajectory of the thin shell.}}
	\label{fig12}
\end{figure}
\begin{figure}[tb]
		\begin{tabular}{ cc }
			\includegraphics[width=3.9cm]{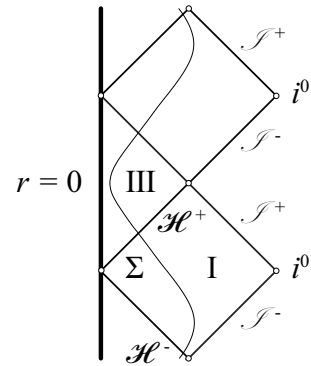}
		\end{tabular}
	\caption{\footnotesize{Penrose diagram of the spacetime for $0 < | Q_{-} + q | = M_{-} + E$ and $\max \{ |q|,E \} < m$. The thin curve is the bound trajectory of the thin shell.}}
	\label{fig13}
\end{figure}
In this Sec.~\ref{sec5}, we consider the case $0 < | Q_{-} + q | = M_{-} + E$.
An extreme RN black hole can be formed via the gravitational collapse of a charged thin shell at a finite advanced time.
The characteristics of the gravitational collapse of the thin shell are same as the aforementioned results of the subextreme case.
Since the formation of an extreme black hole is related to the third law of black hole thermodynamics, we confirm in detail that the advanced time for the formation of the event horizon is finite.
$\dot{T} _{+} \left( \tau \right)$ diverges at the proper time $\tau _{\rm e}$ where $\tau _{\rm e} := R^{-1} \left( r_{\rm e} \right)$.
We take the limit as $\tau \rightarrow \tau _{\rm e}$ of the advanced time $v$:
\begin{eqnarray}
	v \left( \tau \right) = \int_{\tau _{0}}^{\tau} \dot{T}_{+} d\tau + \int_{r_0}^{R\left( \tau \right)} \frac{1}{f_{+}}dr \, .
	\label{eq240}
\end{eqnarray}
To see that the improper integral $v \left( \tau _{\rm e} \right)$ is finite, we check the order of the integrand of $v$.
As $\tau \rightarrow \tau _{\rm e}$, $f_{+}$ approaches zero and $F$ does not.
\begin{eqnarray}
	\frac{\dot{T} _{+}}{\dot{R}} + \frac{1}{f_{+}} = -\frac{1}{2F^2} + O\left( f_{+} \right)\;{\text{ as }}\tau \to \tau _{\rm e} \, .
	\label{eq250}
\end{eqnarray}
Consequently, $v \left( \tau \right)$ can be extended to $\tau _{\rm e}$ in an ingoing Eddington-Finkelstein patch.
The advanced time $v \left( \tau _{\rm e} \right)$ for the formation of the extreme black hole is finite.
The Penrose diagrams of the spacetimes are shown in Figs.~\ref{fig10}--\ref{fig13}.

The third law was first formulated in the following form.
It is impossible by any procedure, no matter how idealized, to reduce the surface gravity to zero by a finite sequence of operations~\cite{Bardeen:1973gs}.
The third law in this form does not, so far, have a proof.

Since it is difficult to define the meaning of a ``finite sequence of operations,'' the following version of the third law is proved and expressed in the formulation.
A nonextremal black hole cannot become extremal (i.e., lose its trapped surfaces) at a finite advanced time in any continuous process in which the stress-energy tensor of accreted matter stays bounded and satisfies the weak energy condition in a neighborhood of the outer apparent horizon~\cite{Israel:1986gqz}.
The third law of this formulation assumes a subextreme black hole as an initial object, so the results in this Sec.~\ref{sec5} are not counterexamples but provides us with interesting examples with superextreme spacetimes as initial objects.
In fact, it can be shown that an extremal black hole can be formed in a finite advanced time even from a Minkowski spacetime by dropping a charged thin dust shell~\footnote{T.H. is grateful to M.~Kimura for pointing out this fact.}.

\section{Concluding Remarks} 
\label{sec6}
It is intuitive that a RN black hole can be charged up by dropping a charged test particle of the same sign onto it, but it is impossible to go further in the extreme case.
This fact has also been confirmed for various higher dimensional black holes, and it is said that ``black holes die hard''~\cite{BouhmadiLopez:2010vc}.

In this paper, we have discussed the formation of a black hole via the gravitational collapse of a charged thin dust shell onto a RN naked singularity.
We have revealed the existence of a transition from a naked singularity to a black hole.
Although, the fluid model fails around the singularity, it is sufficient for the analysis of black hole formation.
Around the singularity, we might have to consider a field model of matter in order to enhance the credibility of the result inside the event horizon. This remains as a task for the future.

Since our system is symmetric under time reversal, there is a reverse process of the motion.
Does a transition from a RN black hole to a RN naked singularity occur by dropping a charged thin dust shell?
It has been asserted that this transition does not occur unless the matter energy density of the thin shell is negative~\cite{Boulware:1973tlq,Gao:2008jy}.

In fact, the real reverse process is a transition from a RN naked singularity to a RN black hole by emitting a charged thin dust shell, and this process is certainly possible.
It should be important to elucidate the relationship between time reversal symmetry in a system and the reversibility of a transition in a spacetime.
It is a way to approach the essence of the weak cosmic censorship conjecture.

Furthermore, in this paper, we saw that an extreme RN black hole could be formed in a finite advanced time by dropping a charged thin dust shell onto a RN naked singularity.
The version of the third law that has been proven~\cite{Israel:1986gqz} is compatible with our result, because it assumes a subextreme black hole as the initial object.
Thus, the weak cosmic censorship conjecture and the third law are closely related, and the validity of the third law is very sensitive to its formulation.

In future work, we would like to consider how generally we can describe in what way the naked singularity is prohibited.
In order to prove that something is prohibited from forming, it is necessary to make clear what kinds of processes are allowed.
From this point of view, it seems worthwhile to consider a more elaborate formulation of the cosmic censorship conjecture based on the findings in the present paper.

\section*{Acknowledgements}
This work was supported in part by JSPS KAKENHI Grant Numbers JP19K03876 (T.H.).
T.H. is very grateful to M.~Kimura and H.~Maeda for helpful discussion.
K.H. would like to thank S.~Kitagawa for constructive comments and K.~i.~Maeda for variable discussion and suggestions.
K.H. is deeply grateful to Y.~Maruyama for encouragement.
K.H. also expresses his gratitude to the people who have supported him.




\begin{thebibliography}{99}

\bibitem{Penrose:1969pc} 
  R.~Penrose,
  Riv.\ Nuovo Cim.\  {\bf 1}, 252 (1969)
  [Gen.\ Rel.\ Grav.\  {\bf 34}, 1141 (2002)].

\bibitem{Wald:1974}
  R.~Wald,
  Ann.\ Phys.\ (N.Y.)\ {\bf 82}, 548 (1974).

\bibitem{Needham:1980fb} 
  T.~Needham,
  Phys.\ Rev.\ D {\bf 22}, 791 (1980).
  doi:10.1103/PhysRevD.22.791

\bibitem{Semiz:1990fm} 
  I.~Semiz,
  Class.\ Quant.\ Grav.\  {\bf 7}, 353 (1990).
  doi:10.1088/0264-9381/7/3/009

\bibitem{deFelice:2001wj}
  F.~de Felice and Y.~Q.~Yu,
  Class.\ Quant.\ Grav.\  {\bf 18}, 1235 (2001).

\bibitem{Hubeny:1998ga} 
  V.~E.~Hubeny,
  Phys.\ Rev.\ D {\bf 59}, 064013 (1999)
  doi:10.1103/PhysRevD.59.064013
  [gr-qc/9808043].

\bibitem{Jacobson:2009kt}
  T.~Jacobson and T.~P.~Sotiriou,
  Phys.\ Rev.\ Lett.\  {\bf 103}, 141101 (2009)
  [Erratum-ibid.\  {\bf 103}, 209903 (2009)]
  [arXiv:0907.4146 [gr-qc]].

\bibitem{Chirco:2010rq}
  G.~Chirco, S.~Liberati and T.~P.~Sotiriou,
  Phys.\ Rev.\  D {\bf 82}, 104015 (2010)
  [arXiv:1006.3655 [gr-qc]].

\bibitem{Hod:2002pm}
  S.~Hod,
  Phys.\ Rev.\  D {\bf 66}, 024016 (2002)
  [arXiv:gr-qc/0205005].

\bibitem{Barausse:2010ka}
  E.~Barausse, V.~Cardoso and G.~Khanna,
  Phys.\ Rev.\ Lett.\  {\bf 105}, 261102 (2010)
  [arXiv:1008.5159 [gr-qc]].

\bibitem{Isoyama:2011ea} 
  S.~Isoyama, N.~Sago and T.~Tanaka,
  Phys.\ Rev.\ D {\bf 84}, 124024 (2011)
  doi:10.1103/PhysRevD.84.124024
  [arXiv:1108.6207 [gr-qc]].

\bibitem{Sorce:2017dst} 
  J.~Sorce and R.~M.~Wald,
  arXiv:1707.05862 [gr-qc].

\bibitem{BouhmadiLopez:2010vc}
  M.~Bouhmadi-Lopez, V.~Cardoso, A.~Nerozzi and J.~V.~Rocha,
  Phys.\ Rev.\  D {\bf 81}, 084051 (2010)
  [arXiv:1003.4295 [gr-qc]].

\bibitem{Matsas:2007bj} 
  G.~E.~A.~Matsas and A.~R.~R.~da Silva,
  Phys.\ Rev.\ Lett.\  {\bf 99}, 181301 (2007)
  doi:10.1103/PhysRevLett.99.181301
  [arXiv:0706.3198 [gr-qc]].

\bibitem{Richartz:2008xm}
  M.~Richartz and A.~Saa,
  Phys.\ Rev.\  D {\bf 78}, 081503 (2008)
  [arXiv:0804.3921 [gr-qc]].

\bibitem{Matsas:2009ww} 
  G.~E.~A.~Matsas, M.~Richartz, A.~Saa, A.~R.~R.~da Silva and D.~A.~T.~Vanzella,
  Phys.\ Rev.\ D {\bf 79}, 101502 (2009)
  doi:10.1103/PhysRevD.79.101502
  [arXiv:0905.1077 [gr-qc]].

\bibitem{Gao:2008jy} 
  S.~Gao and J.~P.~S.~Lemos,
  Int.\ J.\ Mod.\ Phys.\ A {\bf 23}, 2943 (2008)
  doi:10.1142/S0217751X08041402
  [arXiv:0804.0295 [hep-th]].

\bibitem{Lake:1979zzb} 
  K.~Lake,
  Phys.\ Rev.\ D {\bf 19}, 421 (1979).
  doi:10.1103/PhysRevD.19.421

\bibitem{Shapiro:1991zza} 
  S.~L.~Shapiro and S.~A.~Teukolsky,
  Phys.\ Rev.\ Lett.\  {\bf 66}, 994 (1991).
  doi:10.1103/PhysRevLett.66.994

\bibitem{Yoo:2016kzu}
  C.~M.~Yoo, T.~Harada and H.~Okawa,
  Class. Quant. Grav. \textbf{34}, no.10, 105010 (2017)
  doi:10.1088/1361-6382/aa6ad5
  [arXiv:1611.07906 [gr-qc]].

\bibitem{Horowitz:2016ezu}
  G.~T.~Horowitz, J.~E.~Santos and B.~Way,
  Class. Quant. Grav. \textbf{33}, no.19, 195007 (2016)
  doi:10.1088/0264-9381/33/19/195007
  [arXiv:1604.06465 [hep-th]].

\bibitem{Crisford:2017zpi} 
  T.~Crisford and J.~E.~Santos,
  Phys.\ Rev.\ Lett.\  {\bf 118}, no. 18, 181101 (2017)
  doi:10.1103/PhysRevLett.118.181101
  [arXiv:1702.05490 [hep-th]].

\bibitem{Crisford:2017gsb}
  T.~Crisford, G.~T.~Horowitz and J.~E.~Santos,
  Phys. Rev. D \textbf{97}, no.6, 066005 (2018)
  doi:10.1103/PhysRevD.97.066005
  [arXiv:1709.07880 [hep-th]].

\bibitem{Horowitz:2019eum}
  G.~T.~Horowitz and J.~E.~Santos,
  JHEP \textbf{06}, 122 (2019)
  doi:10.1007/JHEP06(2019)122
  [arXiv:1901.11096 [hep-th]].

 \bibitem{Santos:2015iua}
  J.~E.~Santos and B.~Way,
  Phys. Rev. Lett. \textbf{114}, 221101 (2015)
  doi:10.1103/PhysRevLett.114.221101
  [arXiv:1503.00721 [hep-th]].

\bibitem{Figueras:2015hkb}
  P.~Figueras, M.~Kunesch and S.~Tunyasuvunakool,
  Phys. Rev. Lett. \textbf{116}, no.7, 071102 (2016)
  doi:10.1103/PhysRevLett.116.071102
  [arXiv:1512.04532 [hep-th]].

\bibitem{Figueras:2017zwa}
  P.~Figueras, M.~Kunesch, L.~Lehner and S.~Tunyasuvunakool,
  Phys. Rev. Lett. \textbf{118}, no.15, 151103 (2017)
  doi:10.1103/PhysRevLett.118.151103
  [arXiv:1702.01755 [hep-th]].

\bibitem{Gregory:1993vy}
  R.~Gregory and R.~Laflamme,
  Phys. Rev. Lett. \textbf{70}, 2837-2840 (1993)
  doi:10.1103/PhysRevLett.70.2837
  [arXiv:hep-th/9301052 [hep-th]].

\bibitem{Lehner:2010pn}
  L.~Lehner and F.~Pretorius,
  Phys. Rev. Lett. \textbf{105}, 101102 (2010)
  doi:10.1103/PhysRevLett.105.101102
  [arXiv:1006.5960 [hep-th]].

\bibitem{Gibbons:2004au}
G.~W.~Gibbons, S.~A.~Hartnoll and A.~Ishibashi,
Prog. Theor. Phys. \textbf{113}, 963-978 (2005)
doi:10.1143/PTP.113.963
[arXiv:hep-th/0409307 [hep-th]].

\bibitem{Gleiser:2006yz}
R.~J.~Gleiser and G.~Dotti,
Class. Quant. Grav. \textbf{23}, 5063-5078 (2006)
doi:10.1088/0264-9381/23/15/021
[arXiv:gr-qc/0604021 [gr-qc]].

\bibitem{Dotti:2006gc}
G.~Dotti, R.~Gleiser and J.~Pullin,
Phys. Lett. B \textbf{644}, 289-293 (2007)
doi:10.1016/j.physletb.2006.12.004
[arXiv:gr-qc/0607052 [gr-qc]].

\bibitem{Dotti:2008yr}
G.~Dotti, R.~J.~Gleiser, I.~F.~Ranea-Sandoval and H.~Vucetich,
Class. Quant. Grav. \textbf{25}, 245012 (2008)
doi:10.1088/0264-9381/25/24/245012
[arXiv:0805.4306 [gr-qc]].

  \bibitem{Ford:1990ae}
  L.~Ford and T.~A.~Roman,
  Phys. Rev. D \textbf{41}, 3662 (1990)
  doi:10.1103/PhysRevD.41.3662

  \bibitem{Ford:1992ts}
  L.~Ford and T.~Roman,
  Phys. Rev. D \textbf{46}, 1328-1339 (1992)
  doi:10.1103/PhysRevD.46.1328

\bibitem{Nakao:2013uj}
  K.~i.~Nakao, M.~Kimura, M.~Patil and P.~S.~Joshi,
  Phys.\ Rev.\ D {\bf 87} (2013) 104033
  doi:10.1103/PhysRevD.87.104033
  [arXiv:1301.4618 [gr-qc]].

\bibitem{Patil:2011uf}
  M.~Patil, P.~S.~Joshi, M.~Kimura and K.~i.~Nakao,
  Phys.\ Rev.\ D {\bf 86} (2012) 084023
  doi:10.1103/PhysRevD.86.084023
  [arXiv:1108.0288 [gr-qc]].

\bibitem{Lake:1978zz} 
  K.~Lake and R.~C.~Roeder,
  Phys.\ Rev.\ D {\bf 17}, 1935 (1978).
  doi:10.1103/PhysRevD.17.1935

\bibitem{Sato:1986}
  H.~Sato,
  Prog.\ Theor.\ Phys.\ {\bf 76}, 1250 (1986).

\bibitem{Cruz:1967}
  V.~de.~la.~Cruz and W.~Israel,
  Nuovo\ Cimento\ {\bf 51}, 744 (1967).

\bibitem{Kuchar:1968}
  K.~Kucha\v{r},
  Czech.\ J.\ Phys.\ {\bf B18}, 435 (1968).

\bibitem{Chase:1970}
  J.~E.~Chase,
  Nuovo\ Cimento\ {\bf 67B}, 136 (1970).

\bibitem{pontryagin1962ordinary}
  L.~S.~Pontryagin,
  {\it Ordinary Differential Equations},
  (Addison-Wesley, 1962) p. 298.

\bibitem{Bardeen:1973gs} 
  J.~M.~Bardeen, B.~Carter and S.~W.~Hawking,
  Commun.\ Math.\ Phys.\  {\bf 31}, 161 (1973).
  doi:10.1007/BF01645742

\bibitem{Israel:1986gqz} 
  W.~Israel,
  Phys.\ Rev.\ Lett.\  {\bf 57}, no. 4, 397 (1986).
  doi:10.1103/PhysRevLett.57.397

\bibitem{Boulware:1973tlq} 
  D.~G.~Boulware,
  Phys.\ Rev.\ D {\bf 8}, no. 8, 2363 (1973).
  doi:10.1103/PhysRevD.8.2363
    
\end{thebibliography}
\end{document}